\documentclass[runningheads,a4paper]{llncs}

\usepackage{amssymb}
\usepackage{graphicx}
\usepackage{amsmath}

\usepackage{url}
\newcommand{\keywords}[1]{\par\addvspace\baselineskip
\noindent\keywordname\enspace\ignorespaces#1}

\begin{document}

\mainmatter  

\title{Harnessing the power of Social Bookmarking for improving tag-based Recommendations}


%
%
\author{Georgios Pitsilis\inst{1}%
\and Wei Wang\inst{2}\\}
%

\institute{Computer Science Research, Athens, Greece\\
\email{georgios.pitsilis@gmail.com}
\and
School of Computer and Information Technology\\Beijing Jiaotong University, Beijing, China\\
\email{wangwei1@bjtu.edu.cn}}

%
%


\titlerunning{Georgios Pitsilis and Wei Wang}

\maketitle

\begin{abstract}
\emph{Social bookmarking} and tagging has emerged a new era in user
collaboration. \emph{Collaborative Tagging} allows users to annotate
content of their liking, which via the appropriate algorithms can
render useful for the provision of product recommendations. It is the
case today for tag-based algorithms to work complementary to
rating-based recommendation mechanisms to predict the user liking to
various products. In this paper we propose an alternative algorithm
for computing personalized recommendations of products, that uses
exclusively the tags provided by the users. Our approach is based on
the idea of using the semantic similarity of the user-provided tags
for clustering them into groups of similar meaning. Afterwards, some
measurable characteristics of users' \emph{Annotation Competency} are
combined with other metrics, such as user similarity, for computing
predictions. The evaluation on data used from a real-world
collaborative tagging system, \emph{citeUlike}, confirmed that our
approach outperforms the baseline \emph{Vector Space} model, as well as other state of the art algorithms \cite{PengZ09}\cite{PengZeng}, predicting the user liking more accurately.
\end{abstract}


\section{Introduction}

Collaborative tagging, a web-based service that is representive of the
new Web 2.0 technology, allows users to store and share various kinds
of web resources, such as news, blogs, and photos into social data
repositories.
Resources are stored into self-emerging structures called
\emph{folksonomies}, in the form of a post that combines \emph{a)} an
identifier of the resource, \emph{b)} the user who posted it and
\emph{c)} a set of tags. Many web-based resource sharing and
publishing services, like youtube\footnote{www.youtube.com},
flickr\footnote{www.flickr.com}, and Amazon\footnote{www.amazon.com}
have already adopted such model, allowing user-generated tags to
facilitate user information search. The concept of using tags
for on-line annotation of objects, also known as \emph{Social
Bookmarking} or \emph{Collaborative Tagging}, constitutes tags as a
novel source of information. Although the use of tags has been found
very convenient for managing and organizing people's digital material,
from the research perspective it seems to have attracted much interest
in Recommender Systems (RS) in the recent years, with literature rapidly
expanding.

Despite Collaborative Filtering (CF) algorithms being the most adopted
techniques for Recommender Systems, the increasing popularity of
collaborative tagging systems pushed towards to tags being integrated
into the process of recommendation production. Mechanisms which employ
the tags alone for computing item recomendations are less common
\cite{Jaschke07}, while wherever numeric ratings are additionally
provided, they are used complementary to tags for computing item
recommendations \cite{Wei:2011}.
Relying exclusively on the user-provided tags for computing
recommendations, it requires that such information is exploited in the
best way for achieving satisfactory quality of predictions. This is
the case for digital publication services, like \emph{flickr}, and in
general for social networking services, since they provide
no-mechanism for numeric ratings-based evaluation of the published content by
the users.

Different from numeric ratings, tags also carry sementic information
that can be further exploitable. In addition, tag-words can be
classified into hierrarchical ordered systems, called
\emph{taxonomies}, structured upon the natual relationships between
their elements. Measurements like, \emph{Semantic Distance} and
\emph{Relatedness} between tags are computable using the taxonomies.
Knowing such distance can prove very important when needing to group
similar tags together. In some cases grouping can help to overcome
issues like \emph{polysemy} of tags or the use of \emph{synonyms} by
users for annotating the same item. Such issues exist because users
behave differently as far as the way they do annotations, expressing
their own style on this task, that differs from one user to another.

While exploring the information hidden on tags for improving
recommendations has already been a topic for investigation by the
research community in the past, the \emph{Annotation Competency} of users has not been taken into account yet.
In this paper we attempt to utilize the power of taxonomies through
tag clustering, along with giving a useful insight into the \emph{Annotation Competency} of the users.

The rest of the paper is organized as follows: In Section
\ref{sec:RelWorkMot} we explain our motivation and related work
in the field. In Section \ref{sec:PromMod} we reason about the idea of
tag clustering in more detail and describe our algorithm. Section
\ref{sec:Eval} is referred to the evaluation tests we performed and the
results received, and finally in section \ref{sec:ResDisc} we present
our conclusions.
\\

\begin{figure}[htbp]
\centering
\includegraphics[scale=0.6,angle=0]{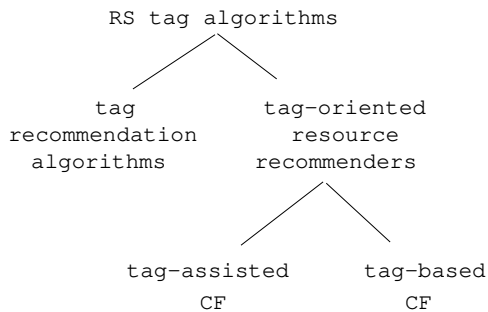}
\caption{Taxonomy of algorithms}
\label{fig:alg}
\end{figure}

\section{Related Work and Motivation}
\label{sec:RelWorkMot}
Based on the existing literature, a simple taxonomy of the tag
recommender systems could be as depicted in Fig.\ref{fig:alg} and it
is explained as follows:
We can distinguish two major types of algorithms, \emph{a)} \emph{tag
recommendation algorithms} and \emph{b)} \emph{tag-oriented resource
recommenders}.
The first category comprizes solutions aiming to ease the process of
annotation by providing personalized recommendations of tags to users
about specific items \cite{Lipczak}\cite{Peng2010}\cite{Symeon2008}.
Mechanisms that belong to the first category can either exist as part
of a larger concept for resource recommendion, or they can stand as
independent services, enabling social network applications to
providing automated annotation of various kinds
\cite{Alipr2008}\cite{Moxley2008}.
The \emph{tag-oriented} category regard prediction models exclusively
for resource recommendations, which can be further divided into two
sub-categories.
For better reference we will call these categories:
\emph{Tag-assisted CF} and \emph{Tag-based CF}.
In \emph{tag-assisted CF} belong those models which require both item
rating values and tags to be available for working out predictions. On
the other hand, \emph{tag-based CF} comprises those models in which
the computation of predictions can be performed using the tags alone.
The first category has been more explored than the second one, hence
the more literature available.
It is interesting to note that almost all proposed models that belong
to this category perceive the task of recommendation production as a
two step process.
First, computing the nessesary similarity correlations, and then
performing the item predictions using the pre-computing similarities
from the first step.

\subsection{Limitations of the Classical Approaches}

There has been a growing number of research efforts that could be
classified as \emph{tag-assisted CF} model. For
instance, in \cite{Liang08} \cite{ZengL08} \cite{Peng10} there has
been an effort to user and tag similarities to be combined together
into a single expression of similarity, while the work by Parra et.al
\cite{ParraB09}, employs a type of tag-based similarity. Finally,
Tso-shutter et.al in \cite{Tso_sutter95} introduce a method of fusing
user-based with item-based CF, treating the user tags as additional
data.

Nevertheless, we focus our interest on \emph{tag-based CF} model
mainly for two reasons. First, because it is very common for tags to
exist as the only available source of information that users provide
in a RS, and second, the less constraints imposed by this model
extends its suitability to a wider range of applications.

For instance, we will refer to one key model from the literature for
\emph{tag-based CF}.
In the work by Peng et al. in \cite{PengZ09}, each tag is viewed as a
distinct topic, while the liking of a user to an item is seen as the
probability of this user to experience that item. The value of this
probability is computed by summing the transition probability through
all tags used for tagging this item. The formula they introduced for
computing the probability $p(i|u)$ that a user $u$ would like item $i$
is given as follows:

\begin{equation}
p(i|u) = \sum\limits_{t \in T} p(t|u) \cdot p(i|t)
\label{eqn:piu}
\end{equation}

where: $T$ is the set of tags used by user $u$, and $p(t|u)$ is the
probability that user $u$ uses tag $t$ for item annotation, and
$p(i|t)$ is the conditional probability of experiencing item $i$ when
tag $t$ is given.

The intuition behind their formula can be phrased as follows: The
liking of a user $u$ to item $i$ is highly related to the probability
that a particular tag is used by that user, as well as the popularity
of this tag when used for annotating item $i$.
That probability $p(i|u)$ is more or less characteristic to the way that a user
makes his own selection of tags for annotating objects. To capture
this for a user we introduce the notion of \emph{Annotation Competency}. We
hypothesize that the quality of predictions received by a user would
certaintly be affected to some degree by his/her such Competency.

We should also point out that the use of Eq.\ref{eqn:piu} becomes
inefficient for the reason that users maintain some own collection of
words they use for annotating. For that reason it becomes less likely
for the probability $p(i|u)$ to be computable, thus strongly affecting the
avaibalility of predictions.
Such formula imposes a serious limitation for the model to
work as it requires a significant overlap to exist between the words
that various users have used. For the reason that it is very much the
case for users' \emph{Annotation Competency} to be as such, we argue that the
above requirement expressed in Eq.\ref{eqn:piu} would result to the
algorithm performing poorly.

In our opinion, a good model should consider as well any differences
that might exist in the \emph{Annotation Competency} from one user to
another.

Another evidence that supports our argument, is the fact that most
systems which belong to the above two \emph{Tag-oriented} categories consider the
relationship between the available sources of data as a tripartite
structure of items-tags-users. Employing such structure has the
weakness of being able to capturing only the two out of the three
binany associations at a time, among the tags, entities and users, yet
something easy to observe and well mentioned by other researchers
 in the field \cite{Peng10}. As such, that would result to sparse
user-tag and item-tag matrices, which in addition to the above
mentioned requirment we set regarding the \emph{Annotation Competency}
would further degrade the recommendation quality.

\subsection{Solutions for Overcoming the Limitations}
\label{sec:OverLimit}

The above remarks suggest the need for taking into account as well the
\emph{Annotation Competency} in the recommendation process. We distinguish
the following two central attributes to describe the \emph{Annotation Competency} of a user.

\begin{itemize}
\setlength\itemsep{0.6em}
\item The \emph{Diversity of Concepts} used by a user throughout
his/her tagging excercise
\item The \emph{Annotation Contribution} on items.
\end{itemize}

The attribute of \emph{Diversity of Concepts} accounts the variety of
topics which characterize the profile of a user's interests.

The intuition behind this attribute is driven from the fact that not all users have the same level of experience nor they show the same willingness in their annotation exercise. We
consider this attribute to be highly important for the quality of
recommendations, mainly because the more experienced or eager users, compared to other users, are meant to be more influencing on the computed predictions. That is because a user who selects tags out
from a corpus that includes words from many disciplines, accually provides
more data for the system training, and hence he should be regarded to
be more contributing than another user whose tags regard only a small
area of topics.

As far as the second attribute, \emph{Annotation Contribution}, while it
captures the same requirement of \emph{Annotation Competency}, as the first one, it though works on item scale.
Concerning \emph{Annotation Contribution}, we perceive that attribute as being
expressed by quantitative criteria and it varies from one user to
another for a particular item. The quantitative aspect in our case interprets as: The more tags provided for annotating an object
the better it is.
That is nessesary for distinguishing the contribution of a user who
has used few, but identical tags, from another user who has used more
descriptive tags for annnotating the same item.
For example the tags 'cat' and 'hungry' add more information to the
context of the annotated subject than if tags 'cat' and 'hungry' were used.
While the first attribute regards the whole reprtoire of tags used
by some user in overall, the second one is refered to only some
particular annotation experience of that user.

Next we present in more detail our design considerations concerning
the two attributes we introduced.

\subsection{Design Considerations concerning the \textit{Diversity of Concepts}}

To implement the first attibute which expresses the Variety in the
Concepts used in a tagging excersize, it requires indentifying all subjects incorprorated into the \emph{Annotation Competency} of a user.

A straightforard approach to identify the concepts is to partition the
tags into distinct \emph{Subjects}. There are models found in the
literature which incorprorate such idea of distincted \emph{Subjects}.
For instance, the work by Peng et al. in \cite{PengZeng}, follows the
concept of organizing the tags into groups. In their work is attempted
a refinement of the technique proposed by the same authors and is
summed in eqn.\ref{eqn:piu}. The refined method employs a type of soft
clustering called \emph{Consistent Nonegative Matrix Factrorization}
(CNMF), that is a method of applying multivariate analysis onto the
tags, for categorizing them into \emph{Subjects}. Their refined
technique is described in the following formula:

\begin{equation}
p(i|u) = \sum\limits_{s \in S} p(s|u) \cdot p(i|s)
\label{eqn:pius}
\end{equation}

where $p(i|u)$ is the predicted liking of user $u$ for item $i$,  $S$
is the set of all identified subjects which tags belong to. $p(s|u)$
is the probability of user's $u$ interest in subject $s$ and $p(i|s)$
is the probability of experiencing item $i$ when users are interested
in subject $s$.
Similarly to their model described in eqn.\ref{eqn:piu}, the refined
one follows the central intuition of social annotation that is: If
user $u$ has used tag $t$ (or a tag of a subject $s$) for many times,
and tag $t$ (or the subject $s$) has been annotated on the article $i$
for many times then it is very likely that user $u$ has strong
interest on the article $i,$ which should finally be recommended to him/her. 
From now on, when referring to this formula we will be using the term \emph{Topic-Based} Variation, for short, while the term \emph{Tag-Based} Variation will be used when refering to their original method in eq.\ref{eqn:piu}

A quite similar approach that works on the same idea of distinuishing
topics of interests from the tags, has also been proposed by Shepitsen
et al. \cite{Shepitsen2008}. Their idea works exactly on the same
principle expressed in eqn.\ref{eqn:pius}, but it employs hierarchical
clustering onto the set of tags for extracting topics.
While we consider the solution of partitioning as a very useful one, nevertheless it does not suffciently capture the users'
\emph{Annotation Competency} in the way we expressed it in the two central
attributes, for making recommendations. For example, to be in line
with the first attribute we set for capturing the diversion in the
areas of interest for a user, if applying partitioning alone would not be enough for achieving this.

Furthermore, for the reason that it is quite common for two different
people to may have chosen different tags for annotating the same item, it is
reasonable to assume that an algorithm that would work on exact
matches on the set of tags used, would not be that efficient for identifying
similar users.
Therefore, the benefit aquired from Collaborative Tagging would not be exploited if using such algorithm.

The logical path to follow would be to group together those
tags which are closer to each other, as far as the contextual meaning
they carry. From then on in the recommendation proccess, tags would be
identified by the cluster ID which they belong to. In this way, it
would suffice to simply knowing the user-to-cluster associations
rather the user-to-item associations, as the existing models require.
That would make easier to spot hidden similarities, and any likely
common interests between users, which otherwise, due to the sparce
user-tag-item relationships, such similarities would not be easily distinguisable.

\subsection{Design Considerations concerning the \textit{Annotation Contribution} of a user}

As we mentioned earlier, the second attribute we indruduced requires
that \emph{Annotation Contribution} has also been described using quantitative criteria.
Next we will refer to the \emph{Vector Space} model \cite{Salton1975},
which is itself a basic tag-based recommendation algorithm, that takes
as input the frequencies of tags for computing the distance or
similarity between users and items.

We mention \emph{Vector Space} model here as a good starting point of
our consideration as it employs qualitative criteria onto the
collection of words used by users for finding potentially interesting
items for them.

Roughly, the vector space model works as follows:
Each user is represented by a tag vector
$u=[w_{t_1},w_{t_2},...,w_{t_n}]$ with $w_t$ denoting the weight of
the patricular tag $t$ on that user. Vector weights may be expressed
through many ways, with the frequency of tags to be the most common.
Likewise, each resource can be modeled as a vector
$r=[u_{t1},u_{t2},...,u_{tn}]$ over the same set of tags. Next, the user's profiles and the resources can be matched over those tag expressions by
computing the similarity value between them.
\emph{Cosine Similarity} can be used to obtain these similarity scores
between user profiles and rated resources. Then, by sorting
the similairites in decending order we eventually get the $Top\_N$ list
of personalized recommendations of resources for a specific user. The
cosine of the two vectors in eqn.\ref{eqn:cosine} is derived from the
Eucledian dot product formula, with $||\vec{t_n}||$ denote as the
length of the vector $t_n$, and $\vec{t_1} \cdot \vec{t_2}$ is the
inner product of the two vectors.

\begin{equation}
cos(\vec{t_1},\vec{t_2})=\frac{\vec{t_1} \cdot \vec{t_2}}{
||\vec{t_1}|| \ ||\vec{t_2}||}
\label{eqn:cosine}
\end{equation}
Since frequency values cannot be negative, cosine similarity will
range from 0 to 1, with 1 denoting a perfect match.
Adapted versions of the vector space model to work with folksonomies
has made this algorihm dominant in the Top-N tag based recomendations
models.

While \emph{Vector Space} model takes into account the individual
annotation scores of users on items for computing personalized
recommendations, in our opinion it only exploits a single dimension of
these data.

Another interesting work which also employs cosine similarity is that
of Xu et al. \cite{Xu2011}. They proposed a \emph{Tag-based} CF
system, which approaches the concept of tag-clustering, in which the
cosine similarity of the frequency of each tag over the set of users
is used to express a form of distance between the tags.
In this way, a resource-tag matrix is composed out of the accumulated
occurience rate of each tag, and is then used as input to a clustering
algorithm. While the motivation of their approach is rational,
finally allocating the similar tags into the same partition, however it does
not take into account the semantic information carried by each tag. We
mention this approach in our survey mainly because it follows a
similar principle for similarity with that of the standard retrieval
model for social tagging systems.

Different from our concept, the work by Gemmell et.al
\cite{Gemmell2008}, while it incoprorates the concept of clustering,
it finally follows the well established principle used in the
\emph{Vector Space} model. In that one, item suggestions are made upon the
computed relevance between users and items.
Similarly to the work by Xu et al.\cite{Xu2011} they perform
clustering on the tags using as input their frequencies of usage in
annotation excersizes.

In our opinion, it would not suffice if using alone the frequencies of
tags in a metric of distance for tag clustering, because,
frequencies can prove not enough to eliminate the implications
of \emph{Synonimy} and \emph{Polysemy}. \emph{Polisemy} exists because
a tag might have multiple related meanings, and \emph{Synonimy} exists
when different tags sharing the same or similar meaning.

We should also mention the existence of works which explored the idea
of studying the relationships between the tags in tag clouds. In the
approach by Venetis et.al in \cite{Venetis11} new metrics were
introduced that capture those relationships. Nevertheless, in these works
the concept of \emph{Annotation Competency} is not approached towards a
recommendation model.

\subsection{Semantic Similarity for Clustering}

The above points highlight the need to take into account the Semantic
Similarity of tags in our model we introduce for personalized
recommendations.

In the way it works, taxonomic similarity for tags works on static
knowledge, which tags are as such by their nature. On the contrary, the
computation of \emph{Cosine Similarity} on the \emph{Vector Space} model,
being based on dynamic data, such as the item-tag data structure,
requires recomputation upon the arrival of new data.
The fact of taxonomic similarity of tags not being dependent on
dynamic data, offers the practical advantage of such metric to work
efficiently with complex algorithms, like Clustering. That means
Clustering should need to run only once, since the distance between
the tags does not change. Instead, in the Vector Space model, the
frequency of each tag does change, as the RS system develops, requiring frequent recomputation of its value.

In a real system, it would suffice to run Clustering just once, during
the system initialization, no matter how much data have been provided
by the users for annotation. Furthermore, cold start issues would be
avoided for the reason that the complete semantic network is established
early on, during system initialization.

We believe that the concept of applying Clustering on the tag corpus,
using the Semantic similarity as a metric for measuring the distance
between the words used as tags, has truly strong potential. To the
best of our knowledge, applying the above concept of Clustering for
improving the performance of \emph{tag-oriented RS}, while taking into
account the users \emph{Annotation Competency}, has not been investigated before.
As far as the concept of Clustering is concerned, it has itself been
the subject of research in RS for either improving the performance of
predictions \cite{Dubois2009}\cite{PitsilisZW11} or for securing RS
against threats, such as profile injection attacks
\cite{Mehta2007}\cite{Su2005}\cite{Zhang2013}.

In our opinion, to achieve a substantial benefit in terms of accuracy
from the appplication of Semantic Clustering, it requires a model that
would capture the characteristics of social tagging systems more
sufficiently and that would incorporate the concept of \emph{Annotation Competency}.
As opposed to existing models, with our prediction model we attempt to
utilize the potential of Collaborative Tagging by fusing our
introduced properties of \emph{Diversity of Concepts} and
\emph{Annotation Contribution} into the concept of \emph{Annotation Competency}.

We sum up the main novelties introduced in our work to the following:

\begin{itemize}
\setlength\itemsep{0.6em}
\item A semantic-similarity based concept for partitioning the user tags.
\item A model for computing personalized item recommendations that
uses as input the \emph{Annotation Competency} of users.
\end{itemize}

Considering the above requirements, in the next section we propose a
model suitable for the \emph{tag-based CF} type.

\section{Proposed Model}
\label{sec:PromMod}

In this section, before we elaborate the details of our concept we
will refer to useful knowledge about the components incorporated in
our design.
 Moreover, we present the design considerations of our approach
derived from the attributes and their requirements we set in the
previous section.

\subsection{Concept Similarity}

In this section we describe in more detail the concept of \emph{Semantic
Distance} we will incorporate in our Similarity model.
For computing the distance between any pair of tags we follow the
intuitive idea of using the semantic similarity in a taxonomy. That
is, the shorter the distance from one tag to another, the more similar
the tags are. In our case, tags are regarded as nodes in the taxonomy
tree. For computing the distance we used a metric introduced by
Resnick \cite{Resnik1995}, which is based on the notion of
\emph{Information Content}. According to this theory, the higher in
the hierrarchy a concept is, the more abstract it is, and hence the
less information it contains. Resnick's metric assumes the association
of probabilities with concepts in a taxonomy and there is also an
IS-A relatioship between them in the hierarchy. In that metric
the similarity between two concepts $c_1$ and $c_2$ is given by:

\begin{equation}
sim(c_1,c_2) = \max_{c \in U(c_1,c_2)} [ - \log p(c) ],
\end{equation}

with function $p: C \rightarrow [0,1]$ such that for any concept $c
\in C$ the value $p(c)$ to be the likelihood of encountering an
instance of $c$. $p$ is monotonic such that, if $c_1$ IS-A
$c_2$ then $p(c_1) \leq p(c_2)$. $U(c_1,c_2)$ denote as the set of
concepts that subsume both $c_1$ and $c_2$ in the hierrarchy. The
actual meaning of that equation is that the more infromation two
concepts share in common, the more similar they are. The information
shared by two concepts is indicated by the information content of the
concepts that subsume them in the taxonomy.

\begin{figure}[htbp]
\centering
\includegraphics[scale=0.65,angle=0]{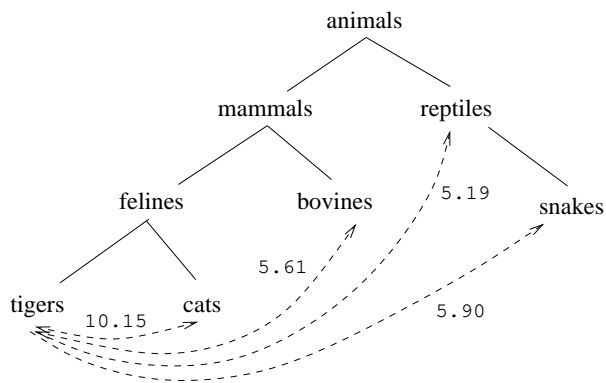}
\caption{Example taxonomy}
\label{fig:taxonomy}
\end{figure}

To enchance clarity and provide a better understanding on how such
model could adapt to the issue we come to address, we give an example.
In the taxonomy of Fig.\ref{fig:taxonomy}, the similarity between
\emph{felines} and \emph{reptiles} equals to similarities between
\emph{tigers} and \emph{snakes}, as well as between \emph{tigers} and
\emph{reptiles} (sim=5.19), for the reason that the set of concepts
that subsumes both of them is \{\emph{animals}\}, and it is common for
them. On the contrary, \emph{tigers} are found to be more similar with
\emph{cats} (sim=10.15), than with \emph{bovines} (sim=5.61).
\emph{Tigers} and \emph{cats} are subsumed by \emph{felines}, that is
a concept of higher information content than \emph{mammals} which
subsumes \emph{tigers} and \emph{bovines}. Therefore, the similarity
of the first pair has higher value than that of the second pair.

\subsection{The Clustering Algorithm}

In order to partition the set of tags into clusters first and formost
we needed a metric to express the distance between a pair of tags.
We considered as distance the similariy values derived from
the application of the Resnick similarity computed onto every pair of tags
from a lexical database. \emph{WorldNet} \cite{fellbaum2005} is
a large lexical database of the English language we chose for our
experimentation. In \emph{WordNet} the various parts of speech are
grouped together into sets of cognitive synonyms, making up a network of
meaningfully related words and concepts.

As far as the clustering algorithm to be used, we chose \emph{Affinity
Propagation} (AP), a newly developed clustering algorithm proposed by
Frey et al. \cite{Frey07}. AP was chosen as it is more efficient than
other conventional approaches, such as \emph{k-means} \cite{kmeans},
and it has shown to achieve remarkably better clustering quality in
various applications. For instance, when AP is applied onto a model for
clustering the users of a collaborative filtering system, it helped to improve the prediction quality \cite{PitsilisZW11}. AP is also known to achieve better performance than if using \emph{K-means} for Abstracting data in anomaly Intrusion Detection Systems (IDS) \cite{Wang2010}. 

\begin{figure}[htbp]
\centering
\includegraphics[scale=0.45, width=165pt, angle=270]{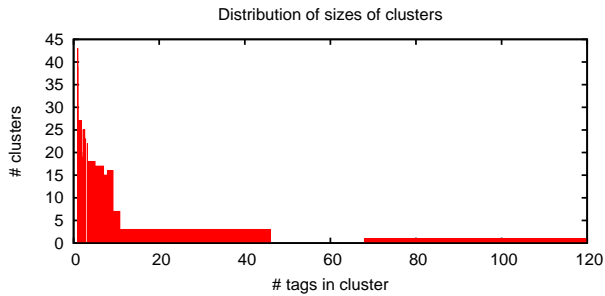}
\caption{Distribution of cluster sizes}
\label{fig:distr}
\end{figure}

Contrary to \emph{k}-means algorithm, in which the number of clusters
is predefined, in AP the quantity of clusters is dependent on the
input data to be clustered, and it can also be affected by a value
called \emph{Preference}. That is a global value applied to each point
expressing its suitability to serve as an exemplar. A big
\emph{Preference} value would cause AP to find many exemplars, while a
small value would lead to a small number of clusters.
Hence, the exact number of clusters emmerges deterministically after a
few iterations of the algorithm.
In our particular case, AP takes as input the similarities of the pairs of tags in the form of $(tag1,tag2,similarity)$ which we considered as the
data points to be clustered.
For initial \emph{Preference} value we chose the minimum similarity
value, as that is the one suggested by the authors of the AP
algorithm.
For our dataset of 3162 tags, after the application of AP, 239
clusters were finally  emmerged, 112 of which had been allocated one
element only. We provide a graph of the distribution of the sizes of
clusters in Fig.\ref{fig:distr}.
Since the internals of tag clustering is out of the scope of our paper
we will not describe the AP algorithm in more detail. The details of the AP can be referred to \cite{Frey07}.

\begin{figure}[htbp]
\centering
\includegraphics[scale=0.60,angle=0]{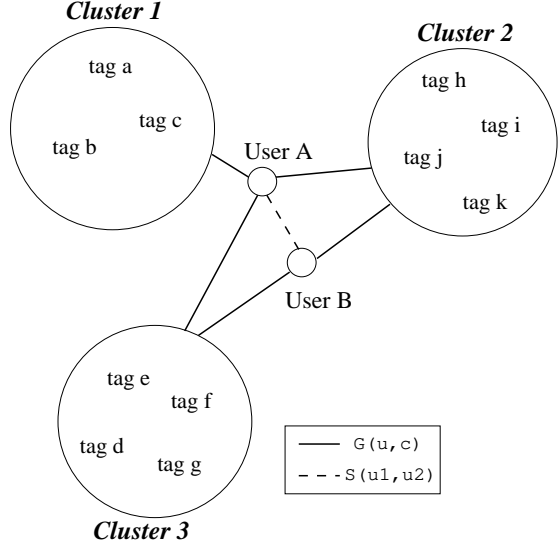}
\caption{Pictorial representation of our concept showing the entities involved.}
\label{fig:concept}
\end{figure}

\subsection{System Model}

Assuming that all tags used by a user would portray his/her personal taste
in annotation, it means that a proper analysis on the tags used by
each user would reveal any hidden similarities that might exist
between a user with others.
Different from the existing approaches mentioned in the previous
section, which mainly exploit the static knowledge provided by the
users, in our proposed model instead there is an attempt to capture
the characteristics of social tagging via the concept of \emph{Annotation Competency} we set in section \ref{sec:OverLimit}.

Examining the problem from a data perspective, the operations on those data can be divided into those applied onto the static ones, like the corpus of tags, and those applied onto the dynamic ones, which are related to the users' excercise on tagging.

In the high level view of our proposed idea that is shown pictorially
in fig.\ref{fig:concept}, can be distinguished, the available tags
being clustered into separate groups, along with the distances between
the users, denoted as similarity value $S(u1,u2)$. In the same
figure, the relatedness of a user $u$ with some cluster $c$ is denoted
as $G(u,c)$.
Considering the above perception on data, Clusters represent the static, while the Sim and G refer to the dynamic part, which is derived from the \emph{user-item-tag} association with the static data.

Next we define new metrics we use for capturing the central attributes of our concept of \emph{Annotation Competency}. The metric of \emph{User Similarity} shown in Def.1 captures the \emph{Diversity of Concepts} expressed by the differences in the tastes of people and in their interests. The \emph{Diversity Value} shown in Def.2 is another metric that captures the \emph{Diversity of Concepts} on the user level. The metric of \emph{Tagging Effort} in Def.3 captures the \emph{Annotation Contribution} of users on particular items. With these 3 metrics combined together can be determined the suitable items to be recommended to a user.
\\

\textbf{Definition 1.}
To capture the central attribute of \emph{Diversity of Concepts} in the tagging excercises on a pair of users $u_i$ and $u_j$ we introduce the
metric of \emph{Similarity}.

Let's call $T$ the set of all tags used by all users. Tags are
partitioned into $N$ clusters $c_1,c_2,...,c_N$, with $C$ the set of
all clusters $C = \{ c_1, c_2, ... , c_N \}$.

We call $E(u_i,t_j,p_k)=\{0,1\}$ a function that specifies whether a
user $u_i$ has tagged item $p_k$ using tag $t_j$.

We call $F$ a function that specifies whether a tag $t_j$ belongs to a
cluster $c_k$ as:

\vspace{3 mm}

$F(t_j,c_k) = \begin{cases} 1 & t_j \in c_k  \\ 0 & otherwise \end{cases}$

\vspace{3 mm}

We introduce a function $G$, we call relatedness and it specifies
whether a user $u_i$ is a member of cluster $c_k$ as:

\vspace{2 mm}

$G(u_i,c_k) = \begin{cases} 1 & \exists t_j,p_k : F(t_j,c_k)=1 ,
E(u_i,t_j,p_k)=1 \\ 0 & otherwise \end{cases}$
\\

\vspace{3 mm}

The value of 1 is received only when the user in question has used at
least one tag that belongs to that particular cluster in his annotation excersize.

Let the set of clusters by user $u_i$ be:

\vspace{3 mm}

$C_{u_i} = \{ c_k | c_k \in C : G(u_i,c_k) = 1 \}, C_{u_i} \subseteq C$

\vspace{3 mm}

To express the similarity between two users $u_i$ and $u_j$ we adapted
a function proposed by Jaccard \cite{jacc01} to the needs of our
concept. The formula of user similarity is given in
Eqn.\ref{eqn:jacc}. Jaccard metric is typically used in the field of
data mining to measure the diversity or similarity in sample sets.

\begin{equation}
S(u_i,u_j) = \frac{|C_{u_i} \cap C_{u_j}|}{|C_{u_i} \cup C_{u_j}|}
\label{eqn:jacc}
\end{equation}

The $||$ indicates cardinality, (i.e., the number of clusters in the set).
$S(u_i,u_j)$ has a range value in the interval $[0,1]$ and maximizes
when the two sets  $C_{u_i}$ and $C_{u_j}$ match. The intuition behind this
formula is that the larger the number of common clusters the tags of the two
users belong to, the more similar the users are, with regard to their taste
in annotation and interests. For example, two users who are both
interested in \emph{cars} and \emph{machinery} are expected to have
used tags which belong to clusters most relevant to \emph{cars} and
\emph{machinery}. Instead, if the first user has used tags that belong
to a cluster that is more relevant to \emph{housing}, it would be
expected to have very low similarity with another user who has used
tags which belong to both clusters of \emph{sports} and
\emph{leasure}.

The central attribute of \emph{Diversity of Concepts} within the
tagging excersize of a single user is captured with the
\emph{Diversity} function we introduce next.
The intuition behind this function is that, users whose interests
comprise many subjects are meant to provide more valuable
contribution, in comparison with other users, as far as the tagging excersize is concerned. Being classified as the most important ones, the opinions of those users will be taken into account for recommending articles to others.
\\

\textbf{Definition 2.} We propose the following function for
\emph{Diversity Value} $w(u_i)$, which returns a binary quantity, by
which we classify whether a user's tagging contribution is valuable or
not.

We define $U_h \subset U$, $U_h=\{u_1,u_2,...,u_k\}$ : $k\in
[1,..,|U|]$, an ordered subset of all users set $U$, such that for any
two users $u_f$,$u_g \in U_h$ and $\forall f,g \in [1,...,k]$, with
$k=|U|$, for which $|C_{u_f}|<|C_{u_g}| \Rightarrow f<g$.
We call $U_{h_1}$ an ordered subset of $U_h$, so that $U_{h_1} = \{ u_1, ... ,
u_h \}$ with $h=\frac{|U|}{2}$. The \emph{Diversity} function returns the
value:

\vspace{1 mm}

$w(u_i) = \begin{cases} 1, & u_i \in U_{h_1}  \\ 0 & otherwise \end{cases}$

\vspace{5 mm}

The binary value received from \emph{Diversity} function $w(u_i)$ is
finally used for filtering out the poorly experienced users, judged on
objective criteria. More particularly, a user that belongs to the top
50\% of the most experienced ones, in terms of diversity in the subjects of interest,  would be considered as a highly contributing user. In our concept, every distinct area of interest is assumed to belonging to a different cluster.

As we mentioned, the above two metrics of \emph{Diversity} in Def.2 and \emph{Similarity} in Def.1, are computed upon both the static and dynamic portion of
data, and therefore their values require recomputation as the user experiences grow.

In order to be in line with the second property of our design, which we called  \emph{Annotation Contribution}, we adopt the following intuition: We
consider those users who have put more effort in annotating some
particular items, as being the strongest candidates for
recommeding these items to other users. To be consistent with our desing principals, the prediction mechanism should be more sensitive to the quantity and the
diversity of tags used by some user for annotating a particular item. Therefore,
we find nessasary to introduce the notion of \emph{Tagging Effort}, that we express here in the form of a metric and we use it as a complementary criterion for filtering out
the items of lower interest from being recommended to users.

\vspace{3 mm}

\textbf{Definition 3.}
We define the metric of \emph{Tagging Effort}, $f$ on some item $p_k$
as follows.
We call $T$ the set of all tags used by all users and $T_{u_i} \in T$
the set of tags used by a particular user in his annotation
excersize. We call $T_{(u_i,p_k)} \in T_{u_i}$ the subset of tags used by
$u_i$ for tagging item $p_k$.
Tagging Effort $f(u_i,p_k)=\frac{|T_{(u_i,p_k)}|}{|T_{u_i}|}$ is
defined as the fraction of tags used by user $u_i$ for tagging the
candidate item $p_k$ over all tags used by that user.

To investigate the level of contribution for each of the two criteria of \emph{Diversity of Concepts} and \emph{Annotation Contribution} into the quality of predictions, we introduce the contribution factor $d$.
We use the following formula to combine together the above two criteria expressed as per metrics of \emph{User Similarity} and \emph{Tagging Effort}. The probability of an item $p_k$ to be recommended to
user $u_i$ by another user $u_j$ is computed as:

\begin{equation}
p(u_i,u_j,p_x) =  d \cdot S(u_i,u_j) + (1-d) \cdot f(u_j,p_k)
\label{eqn:prob}
\end{equation}

Finally, the probability of an item $p_k$ to be liked by user $u_i$ is
given in equation \ref{eqn:liking}, and it represents the normalized liking of
the particular item over all $m$ users who have also experienced the
item $p_k$.

\begin{equation}
p(u_i,p_x) = \frac{ \sum\limits_{j=1}^m [ p(u_i,u_j,p_x) \cdot w(u_j) ] }{m}
\label{eqn:liking}
\end{equation}

\section{Evaluation}
\label{sec:Eval}

For computing the similarity between the tags of users we used
\emph{Wordnet::Similarity}, a freely available software package by
Pedersen et al. \cite{Pedersen2004}, that is written in perl.
This package provides various measures of relatedness including
Resnick's metric which we finally chose to use in our experiment. \emph{WordNet::Similarity} implements the similarity proposal for IS-A
relationships in \cite{Resnik1995}.

We chose the \emph{CiteUlike} dataset as the most appropriate set for our
evaluation.
\emph{CiteUlike} is a public social bookmarking site aiming to promote
and develop the sharing of scientific references amongst researchers.
One can add a scientific references and then add tags of his choice,
allowing to other users to search for references by keywords.
The data we used was taken from an available snapshot retrieved in
2009 from the \emph{CiteUlike} website, and that is provided for
research purposes \cite{citeUlike}. This dataset is available in the
form of a single file, every line of which is consisted of four
elements: \emph{a)} the id of an article annotated, \emph{b)} the ID
of the user who annotated the article and \emph{c)} the tag word used
for annotation, and \emph{d)} the time of the annotation. From the
above fields we can easily build the associations between
\emph{users}, \emph{tags} and \emph{articles}.
For the needs of our experiment, and due to the fact that the original
dataset was very large and sparse, we finally chose  a subset of 1000
users, randomly selected out of the 46444 users contained in the
original set.

For computational efficiency and recommendation quality we applied
filtering onto the selected articles so that only those which have
been annotated by at least 15 users were finaly considered in the
evaluation. In addition, articles that had been annotated for more
than 75 times were excluded. For the same reason we applied filtering
on the tags too, considering only those that have been used for at
least 10 times in the training set. Finally, we also applied filtering
on the users set. Thus, the final 1000 user dataset we used, contained only
users who had annotated at least 20 articles. We chose these filtering
values in order to minimize the impact of the use of reduced dataset
on the tested algorithms.

We performed 5-fold cross validation over the 1000 user data set to
test all algorithms. That is, we randomly divided the user dataset
into 5 subsets of 200 users each, where in each fold we kept the
annotations of one subset of users hidden and tried to predict the
liking of those users, using the remaining 4 folds. The former and the latter subsets are known as \emph{test set} and \emph{training set}
respectivelly. In the prediction phase we used our algorithm to recommend the \emph{top} 20
articles for each user, and compare them with the
actual articles found in the \emph{test set} for the same user.

To distinguish the probable articles we set a \emph{Threshold
Probability} value of zero as the probability of an article must not be equal to, in order to be counted as a probable article overall.
Then, for every user we compiled a \emph{top} 20 list which includes
those articles whose predicted probabilities to be liked by that user
would have exceeded the \emph{Threshold Probability} value.
We call a \emph{hit} an article which has been selected in the
\emph{top} 20 list of a user and for which it trully happens to be one of the items which the user has annotated. We assume that users only annotate items which they trully like.
We measure the number of correct recommendations, or \emph{hits} with
the symbol \emph{$N_{hit}$}.
\emph{$N_{rec}$} is the total number of recommendations and it counts
those cases in which the computed probability $p(u_i,p_x)$ in eq. \ref{eqn:liking} has
received a positive value. \emph{$N_{test}$} is the number of articles
in the test set. After applying the filtering there were in total 845
articles found to meet the criteria, which composed our test set
($N_{test}=845$). More details about the data used can be found in
table \ref{tab:dataDescr}.

\begin{table}[!htbp]
   \centering
   \caption{Data description table. *A \emph{transaction} indicates an
instance of a single tag out of all tags used by some user for
annotating an article.}
   \label{tab:dataDescr}
   \begin{tabular}{|c|c|}
   \hline
    Metric & Value \\
   \hline
   \hline
    Number of users after applying filtering & 518  \\
   \hline
    Number of articles after applying filtering & 845 \\
   \hline
    Number of total tags used by filtered users & 3162 \\
   \hline
    Number of transactions$^*$ & 4931 \\
   \hline
    Avg. Number of articles per user & 1.631 \\
   \hline
    Avgerage frequency of selected tags & 3.062 \\
   \hline
    Number of clusters of Tags & 239 \\
   \hline
   \end{tabular}

\end{table}

Moreover, we performed further analysis onto the results for the
proposed algorithm, measuring the performance for the
\emph{highly} active and the \emph{least} active users in separate.
For classifying the \emph{highly} active users we used the \emph{Median} of the number of clusters $|C_{u_i}|$ of each
individual user as a threshold value. As defined earlier in Def.1,   $C_{u_i}$ is the the number of
clusters which the tags of user $u_i$ are spanning to, meaning that a
user with  $|C_{u_i}|$ value larger than the chosen threshold would be
considered as a highly contributing user and hence as a highly
active one.

For measuring the ability of our proposed algorithm to provide a list
of recommendations of articles that users actually like, we used the
evaluation method called \emph{Precision and Recall}. This method
measures this ability in terms of \emph{Classification Accuracy} and
it is widely used in Information Retrieval
\cite{Herlocker04}\cite{McNee2006}. The
metrics used in Classification accuracy are \emph{Precision} (P), \emph{Recall} (R) and
\emph{F\_Score} (F). For the case of systems that generate \emph{Top\_N}
recommedations, like ours, the definitions of \emph{Precision} and \emph{Recall} are
slightly adjusted from the standard way used in Information Retrieval.

\emph{Precision} indicates the success of the algorithm regarding
whether some recommendation provided by the algorithm for some
particular user matches a real liking of that user. \emph{Precision}
is defined as the ratio of $\frac{size \; of \; hit \; set}{size \; of
\; top\_N \; set}$. The relative success in retrieving all items liked
by individual users is expressed with \emph{Recall}.
Finally, the trade-off between P and R is measured with the
\emph{F\_score}, which is the harmonic mean of the two values. The
metrics used are shown in table \ref{tab:Metrics}.

\begin{table}[!htbp]
   \centering
   \caption{Evaluation Metrics}
   \label{tab:Metrics}
   \begin{tabular}{|c|c|}
   \hline
   Metric & Formula used \\
   \hline
   \hline
   Precision & $\frac{size \; of \; hit \; set}{size \; of \; top\_N \;
set} = \frac{N_{hit}}{N_{rec}}$ \\
   \hline
   Recall & $\frac{size \;of \; hit \; set}{size \; of \; test \; set}
= \frac{N_{hit}}{N_{test}}$ \\
   \hline
   F.score & $\frac{2 \cdot Precision \cdot Recall}{Precision + Recall}$ \\
   \hline

   \hline
   \end{tabular}

\end{table}

For reference we also tested the classical \emph{Vector Space} method onto
the same data, which employs no clustering. To perform this we
computed the cosine similarity as in eqn.\ref{eqn:cosine} between every pair of
users and items and we finally selected a list of Top\_20 most similar
items to recommend for each user.
We also compared against the while simple, but powerful alternative
\emph{Tag-Based} recommendation method by Peng et al., expressed in eqn.
\ref{eqn:piu}.
For comparison and for showing whether the use of the
transition probability over all subjects might work better with
clustering, we also evaluated the \emph{Topic-Based} variation by
Peng et al., expressed in eqn.\ref{eqn:pius}. Finally, in our evaluation we used our clustering approach for deriving the various subject categories.

To reason whether prediction schemes do actually worth over doing
selections without definite aim, we also tested our method against a
random selection scheme. The main idea of random selection scheme is to build up
the lists of recommended items for each user, in which the top items (20 in our experiment) will be
randomly selected out of the $N_{test}$ items of the whole set. Next,
the number of correct recommendations $N_{hit}$ is counted for each
user as normal.

\section{Results - Discussion}
\label{sec:ResDisc}

\begin{figure}
\centering
\includegraphics[scale=0.44,angle=270]{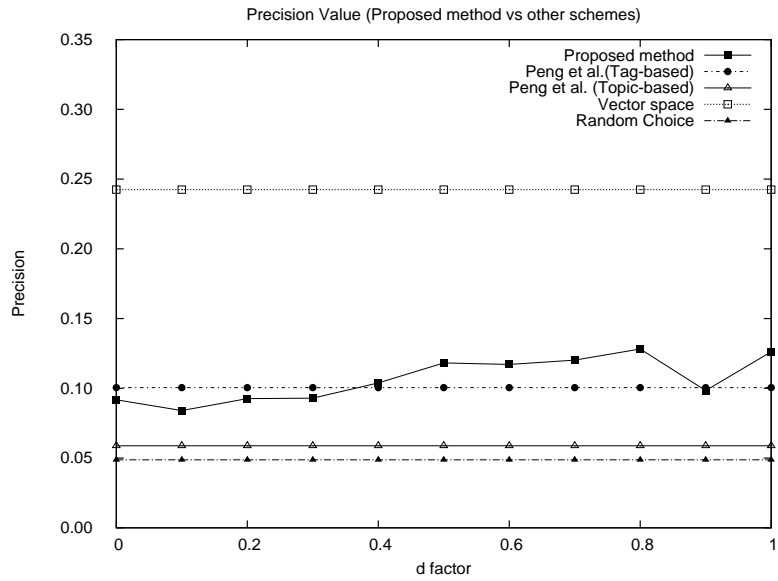}
\caption{Comparison of the proposed method against other algorithms in
terms of Precision}
\label{fig:P}
\end{figure}

\begin{figure}
\centering
\includegraphics[scale=0.44,angle=270]{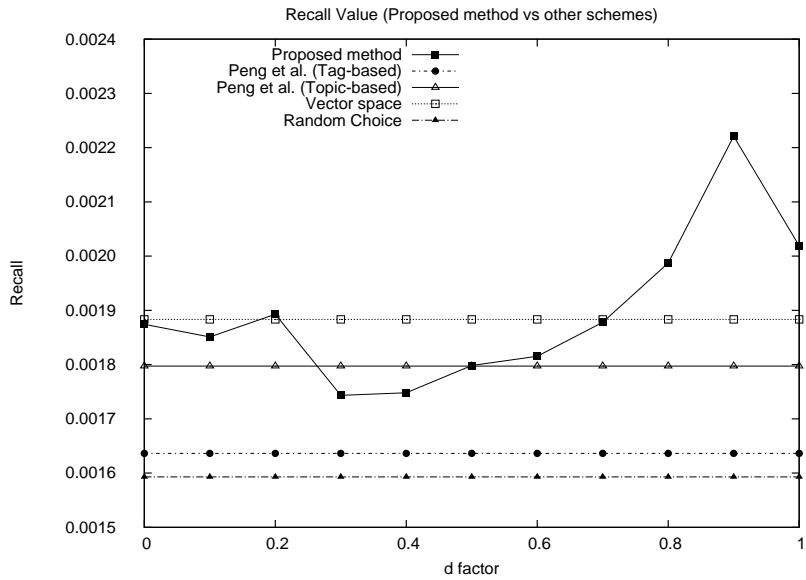}
\caption{Comparison of the proposed method against other algorithms in
terms of Recall}
\label{fig:R}
\end{figure}

\begin{figure}
\centering
\includegraphics[scale=0.44,angle=270]{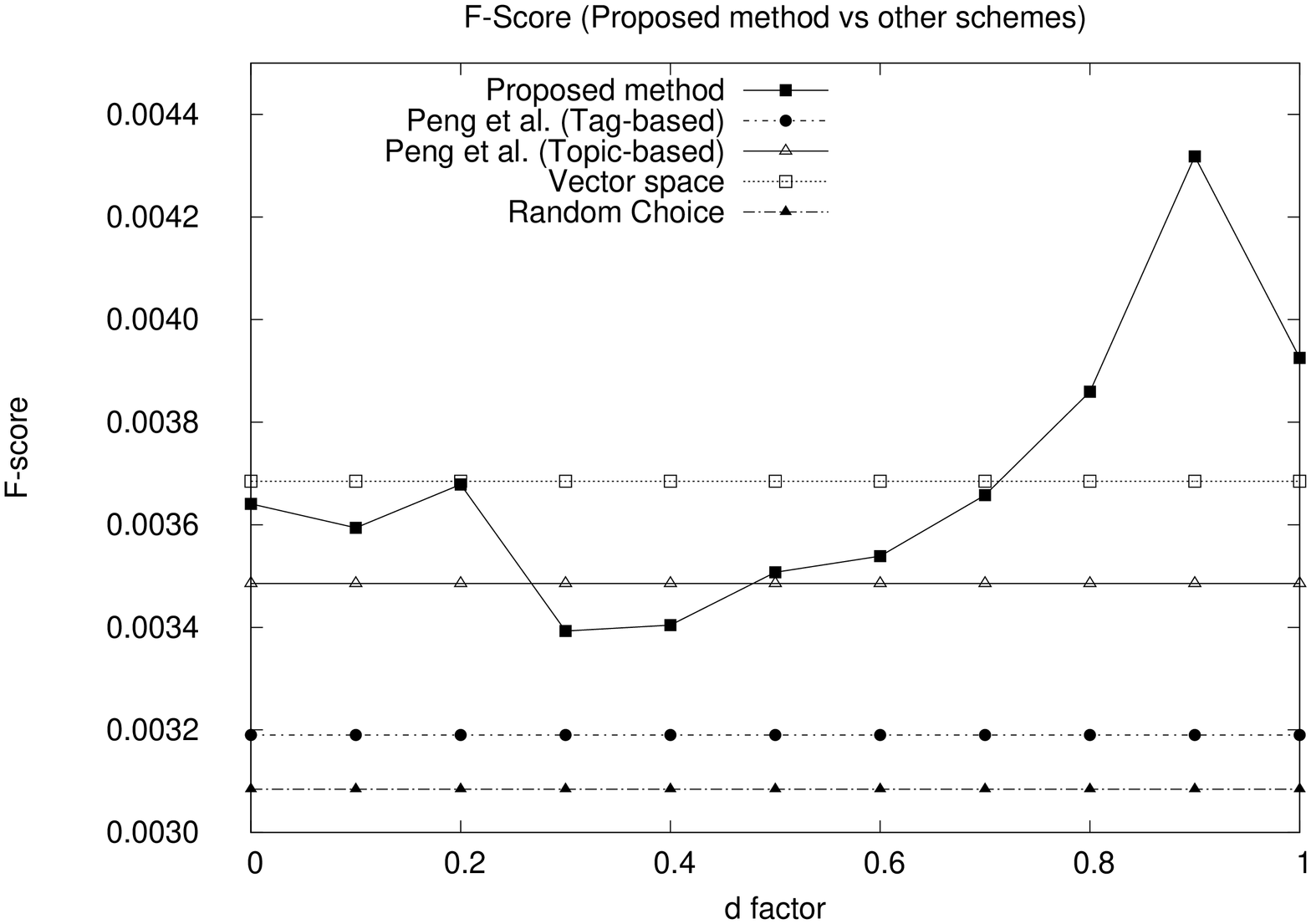}
\caption{Comparison of the proposed method against other algorithms in
terms of F\_score}
\label{fig:Performance}
\end{figure}

We report the most interesting results of our experimentation. The
data presented in table \ref{tab:Results} shows the average values of
10 measurements. The largest values of \emph{P},\emph{R} and \emph{F\_score} are highlighted
in bold. We tested our scheme for various values of the $d$ factor,
ranging from 0.0 to 1.0. This was mainly done to study whether a
mixture of the two criteria has any effect on the prediction quality. We also
include in our report the performance figures of \emph{Tag-based Recommendation}
method by Peng et.al \cite{PengZ09}, and its variation we called \emph{Topic-based Recommendation} \cite{PengZeng}, that we compare ours against to.
All comparative results are shown pictorially in fig.\ref{fig:P},\ref{fig:R} and
\ref{fig:Performance}.

\begin{table*}
   \centering
   \caption{Numeric Results}
   \label{tab:Results}
   \begin{tabular}{|c|c|c|c|c|c|}
   \hline
   Method & R & P & F & F (most active) & F (least active)  \\
   \hline
   \hline
   \emph{Random Choice}            & 0.001593  & 0.04874 & 0.003084 & - & - \\
   \hline
   \emph{Peng et al.}              & 0.001636  & 0.10046 & 0.003190 & - & - \\
   \hline
   \emph{Peng et al. (topic-based)} & 0.001797    & 0.05881   &
0.003485 & -  & - \\
   \hline
   \emph{Vector Space.} & 0.001883  & \bfseries{0.24242} & 0.003685 &  - & - \\
   \hline
   \emph{proposed},  $d=0.0$ & 0.001874 & 0.09176 & 0.003641  &
0.003967  & 0.003325 \\
   \hline
   \emph{proposed},  $d=0.1$ & 0.001851 & 0.08408 & 0.003594  &
0.003939  & 0.003262 \\
   \hline
   \emph{proposed},  $d=0.2$ & 0.001892 & 0.09261 & 0.003678  &
0.003976  & 0.003395 \\
   \hline
   \emph{proposed},  $d=0.3$ & 0.001743 & 0.09295 & 0.003393  &
0.003733  & 0.003070 \\
   \hline
   \emph{proposed},  $d=0.4$ & 0.001748 & 0.10385 & 0.003404  &
0.003728  & 0.003102 \\
   \hline
   \emph{proposed},  $d=0.5$ & 0.001798 & 0.11824 & 0.003404  &
0.003965  & 0.003086 \\
   \hline
   \emph{proposed},  $d=0.6$ & 0.001815 & 0.11714 & 0.003539  &
0.003829  & 0.003275 \\
   \hline
   \emph{proposed},  $d=0.7$ & 0.001878 & 0.12024 & 0.003658  &
0.003919  & 0.003412 \\
   \hline
   \emph{proposed},  $d=0.8$ & 0.001987 & 0.12815 & 0.003859  &
0.004099  & 0.003650 \\
   \hline
   \emph{proposed},  $d=0.9$ & \bfseries{0.002221} & 0.09837 &
\bfseries{0.004318}  &  \bfseries{0.004577}  & \bfseries{0.004076} \\
   \hline
   \emph{proposed},  $d=1.0$ & 0.002019 & 0.12618 & 0.003925 &
0.003775    & 0.004070 \\
   \hline
   \end{tabular}
\end{table*}

We make the following observations on the results.
As can be seen in figures \ref{fig:P},\ref{fig:R} and \ref{fig:Performance}, our
method outperforms all the alternative algorithms we compared
against it.
More specifically, our method performs best when the \emph{d} factor receives extreme values (\emph{b} $\rightarrow$ 0 or \emph{b} $\rightarrow$ 1).
Moreover, the proposed method outperforms all the other alternatives in terms of
\emph{F\_Score}, when $d>0.7$, with performance reaching its
peek for $d=0.9$. According to our results, the \emph{Vector
Space} model is the second best performing alternative, with the third best to be the \emph{Topic-based} recommendation method by Peng et.al., which our method outperforms for almost all values of
\emph{d}, (except for $d=0.3$ and $d=0.4$).
The observed peak value for \emph{F\_score} at $d=0.9$ can be interpreted as saying: in our proposed method the \emph{Tagging Effort} criterion on particular
items is less significant than that of \emph{Profile Similarity}.

More precisely, at that peak value of \emph{F\_Score\emph}, our method ($d=0.9$) appears to be 17.17\% better than the second best (\emph{Vector Space}
model) achieving \emph{F\_score}=0.00432 vs 0.00368.

From the diagrams of \emph{P} and \emph{R} in fig.\ref{fig:P} and \ref{fig:R}, we observe that the good performance of our method is terms of \emph{F}, in relation to the other methods we compared it with, is due to the high \emph{R} values achieved.
If considered the \emph{Precision} values alone, the \emph{Vector Space} model
would have been the best performed. On the contrary, in terms of \emph{Precision},
the value of \emph{d} does not appear to have the same strong impact on
the performance as it does for \emph{Recall}.

Moreover, looking at the diagrams of \emph{Precision} and \emph{Recall} more
carefully, we can observe a significant drop in the \emph{Precision} values for a long range of the $d$ factor. Nevertheless, this drop seems to be not enough to eradicate the advantage of \emph{Recall} values in our method, which achieved for large values of \emph{d}. This observation can be interpreded as saying that the \emph{Profile Similarity} criteria is more important for achieving good predictions. On the contrary, in terms of \emph{Precision}, the value of $d$ does not have a strong impact on the performance.

Comparing against the method by Peng et.al alone, we conlcude that,in overall, their both variations produced significantly lower figures of performance than ours, in all aspects.

\begin{figure}
\centering
\includegraphics[scale=0.44,angle=270]{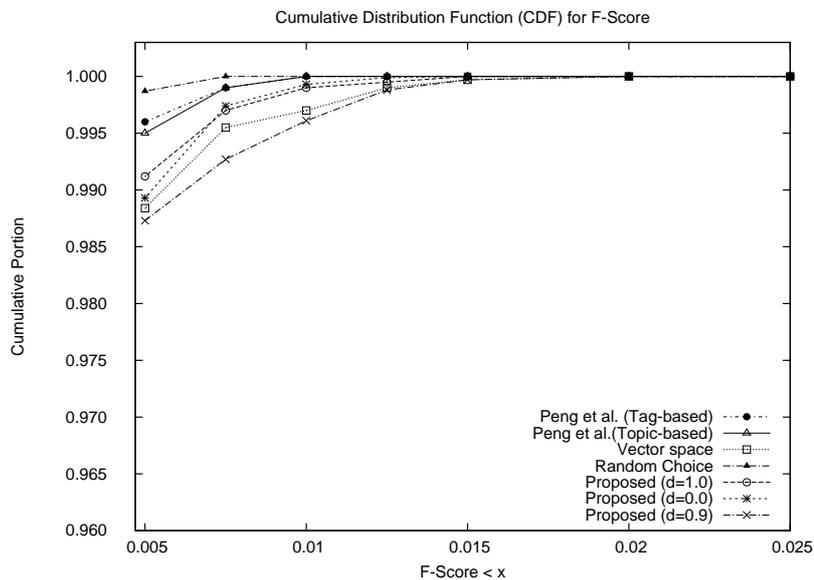}
\caption{CDF of \emph{F.Score} in the proposed method against other algorithms}
\label{fig:CDF}
\end{figure}

To investigate the distribution of the \emph{F\_Score} in the number of users, we
also demonstrate the Cumulative Distribution Function (CDF) of \emph{F\_Score} in figure
\ref{fig:CDF}. CDF describes the probability that \emph{F\_Score} receives a value less or equal to x   ( $Pr(\emph{F\_score}\le x)$ ). In a good
model, \emph{F\_score} would receive as large as possible values, meaning that
the CDF curve should go up as less quickly as possible for a good
model, meaning that a curve close to the right-bottom corner of the diagram
indicates a good model.
In general, a CDF curve that is away from the left-top corner of the
diagram indicates a good model.
In figure \ref{fig:CDF} we present the CDF of the \emph{F\_Score} of our proposed method, as well as the two variations of the Peng et al.
technique, and the \emph{Vector Space} model. We also include the performnce
of the Random selection scheme for recommending items.

In total we test our proposed technique for 3 different values of the
$d$ factor, 0.0, 0.9 and 1.0. The choice of values for the $d$ factor was done using the following reasoning: 1.0 and 0.0 were chosen as the extreme
values which indicate the sole application of either of the two criteria of
\emph{Tagging Effort} or \emph{Profile Similarity} The value 0.9 was chosen as an intermediate case in which the method behaves best in terms of Classification Accuracy.
We observe that our proposed method performed best in terms of CDF, only for the case that a mixture of criteria was applied. More particularly, the best results achieved for $d=0.9$.

Compared to the alternative methods we included in our
evaluation, we observe that our proposed technique has shown the best
behaviour, with \emph{Vector Space} and \emph{Topic-Based}
model by Peng et. al to have achieved poor results. Given a scenario
that \emph{F\_Score} would not fall below than 0.0075, in our
proposed technique only the 99.27\% (for $d=0.9$) of the users would
behave as such. That is to say: the \emph{F\_Score} has 99.27\% chance to not
exceed the value of 0.0075, while for the vector space model, the chance of exceeding
the same value is even larger (99.55\%). In the method by Peng et al, as well as in their \emph{Topic-Based} variation that makes use of subjects, the chance is also higher, (99.9\% for both). For the random policy model, that chance reaches the 100.0\%. Compared to each other in therms of CDF, our
technique in overall (Considering all cases where \emph{F\_score$ < x$}) produces better results, by just 0.126\%  than the vector space model, 0.412\% than the peng et al. method, and 0.389\%  than the subjects version of the algorithm by the same authors.

Despite the marginal superiority of our method, we can intrepret the promishing results of our approach as saying: In
our method it is more likely for the \emph{F\_Score} to receive higher value
than in any other model used in our experiment.

\begin{figure}
\centering
\includegraphics[scale=0.44,angle=270]{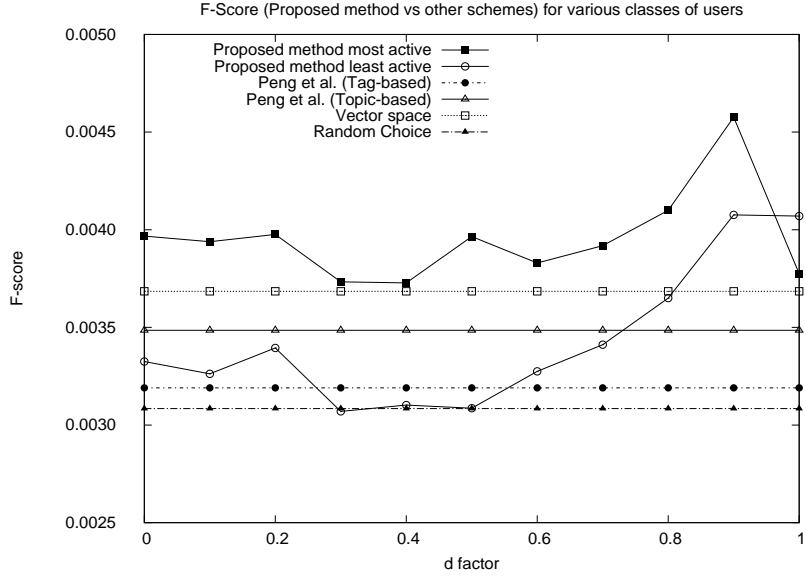}
\caption{Comparison of the proposed method against other algorithms
for various classes of users}
\label{fig:F_most_least}
\end{figure}

We also present separate performance figures in table
\ref{tab:Results} and fig. \ref{fig:F_most_least} for specific
classes of users, like the \emph{Most active} and the \emph{Least
active} ones.

For the class of \emph{Most Active} users our method does significalty better than for the mixed population and it outperforms all the approaches it was tested against in terms of \emph{F\_Score} in the whole range of \emph{d} values. As can be seen
the best performance is achieved when a mixture of
criteria is applied ($d=0.9$), achieving 19.50\% better accuracy than the second best approach, (\emph{Vector Space} model). It is interesting to note that this value is the highest ever counted for all categories of users we tested (\emph{Most active}, \emph{Least active} and the mixed group of users).
On the contrary, for the \emph{Least actrive} users, the performance of our method produced a lower figure, but it still outperformed the \emph{Vector Space} for $d>0.8$. More
specifically \emph{F\_score} ranged from performance levels as low as that of the random choice (\emph{d}=[0.3,...,0.5], F=0.003084), but finally achieving the best performance (F=0.004070) for that category of users for $d=1.0$.

\begin{figure}
\centering
\includegraphics[scale=0.44,angle=270]{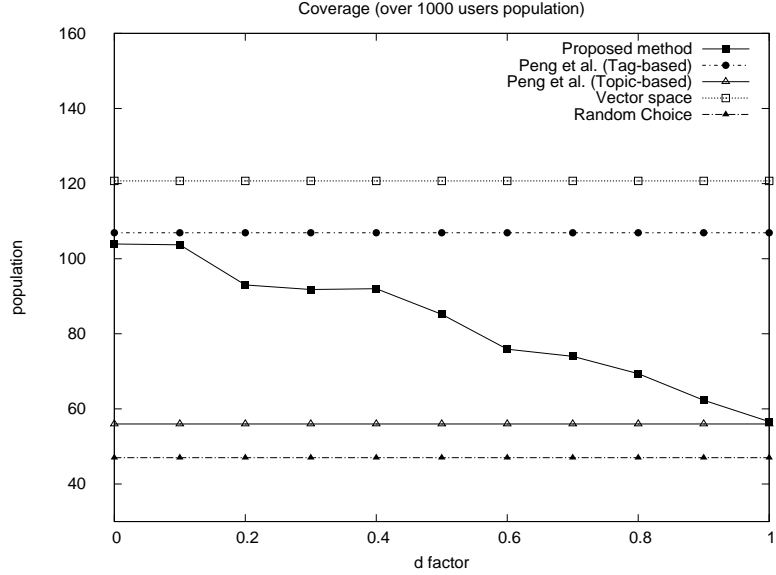}
\caption{Comparison of the proposed method against other algorithms in
terms of Coverage}
\label{fig:Cov}
\end{figure}

Beside Classification Accuracy, it is equally important to know the
success of the method as far as the population of users that can actually
receive the recommendation service.
With \emph{Covered Population} we refer to the users who were able to
receive at least one recommendation for their articles included
in the test set.
Data sparsity is the reason that not all 1000 users from the sample could finally receive recommendations. We present the results of
\emph{Covered Population} for all algorithms we compared against in table
\ref{tab:Coverage} and in fig.\ref{fig:Cov}.
As can be seen the \emph{Covered Population} is indeed affected by the
use of clusters. On the contrary, the \emph{Topic-based} variation of the
algorithm by \emph{Peng et al.} is affected the most, allowing only to
the 5.6\% of the total population (56 over 1000 users) to receive
recommendations.
Instead the \emph{Vector Space} method is the least affected in terms
of coverage.
As we can observe, our method is becoming more sensitive with the increase of \emph{d}, achieving \emph{Coverage Population} that ranges from 103.9 ($d=0$) to 56.6 ($d=1$).
At the peak performance in terms of \emph{F\_Score} ($d=0.9$), the \emph{Coverage} of our method shows a significant drop, as opposed to the \emph{Vector Space} model, with the number of users who can accually receive recommendations to have been reduced in half (62.3 vs 120.7).

\begin{table}
   \centering
   \caption{Coverage over 1000 users population}
   \label{tab:Coverage}
   \begin{tabular}{|c|c|}
   \hline
   Method               & Covered     \\
                        & Population  \\
   \hline
   \hline
   Random Choice        &  47.0   \\
   \hline
   \emph{Peng el al.}   &  106.9  \\
   \hline
   \emph{Peng et al. (Topic-based)}  &  56.0 \\
   \hline
   \emph{Vector Space} & 120.7 \\
   \hline
   \emph{proposed},  $d=0.0$ & 103.9  \\
   \hline
   \emph{proposed},  $d=0.1$ & 103.7  \\
   \hline
   \emph{proposed},  $d=0.2$ & 93.0  \\
   \hline
   \emph{proposed},  $d=0.3$ & 91.8  \\
   \hline
   \emph{proposed},  $d=0.4$ & 92.0  \\
   \hline
   \emph{proposed},  $d=0.5$ & 85.2  \\
   \hline
   \emph{proposed},  $d=0.6$ & 75.9  \\
   \hline
   \emph{proposed},  $d=0.7$ & 74.0  \\
   \hline
   \emph{proposed},  $d=0.8$ & 69.4  \\
   \hline
   \emph{proposed},  $d=0.9$ & 62.3  \\
   \hline
   \emph{proposed},  $d=1.0$ & 56.6  \\
   \hline
   \end{tabular}
\end{table}

In addition, the figures show that the clustering algorithm we chose (\emph{Affinity
Propagation}) applied partitioning onto the tags in a way that
resulted to receiving a large number of small clusters, making the computation of Similarity betweed users less probable.
That posed a serious implication on the number of the computed recommendations.
As such, the chance for the topics of users' interests to overlap is reduced. For that reason we conclude that clustering with \emph{Affinity Propagation} is not
suitable for the \emph{Topic-based} variation method by Peng et al. (see
eqn.\ref{eqn:pius}), while for our method it worked
beneficially for a whole range of values of \emph{d}. For example, for
$d \ge 0.5$ our method does better both in terms of \emph{Coverage} and
\emph{F\_Score}, against the \emph{Topic-based} by Peng et. al.
The higher performance achieved by our technique, which also employs
partitioning for distinguishing the subjects used, justifies the importance of the
notions of \emph{Diversity of Concepts} and \emph{Annotation Contribution} we introduced.

The computational cost of our approach to generate recommendations for
a user comes from the participation of the following 3 factors:

\begin{itemize}
\setlength\itemsep{0.6em}
\item User Similarity computation - (as per eqn.\ref{eqn:jacc})
\item Tagging Effort of a user to particular item - given by $f(u_i,p_k)$ (as per Def.3)
\item Probability for an item to be liked by a user - (as per
eqn.\ref{eqn:liking}).
\end{itemize}

We used the following notation, with $m$ denoting the users, $n$ the items, $l$
the tags and $c$ the clusters. As such, the time complexity of the similarity
computation is $O(c(m+m))$. Similarly, the time complexity of the
Tagging effort is $O(l+nl)$ and the complexity of computing the
probability of liking for an item is $O(m)$. In overall, the complexity of our
model is $O(m[2c(m+m)+l+nl])$ = $O(m(cm +nl))$ = $O(m^2c + mnl)$.
Respectivelly, for the conventional method (\emph{Tag-based}) by Peng et al. expressed in eqn.\ref{eqn:piu}, the complexity is $O(l(n+nl+m+ml))$ =
$O(l^2n+l^2m+ln+lm)$. As can be seen, the second parts in the two
expressions (ours and \emph{Tag-Based} Pend et al.) denote complexity of
equal degree (3rd). The complexity of \emph{Topic-based} variation of Peng et al.
method is $O(c(n+nc+m+mc))$ = $O(c^n+c^2m+cn+cm)$, and it is
nearly equal to that of the first variation, if not including the cost
of clustering.

Likewise, the complexity of the vector space method is computed as follows: The time compexity of the task of computing the tag frequency tables over
all users and all articles are: $O(lm)$ and $O(ln)$ respectivelly. The
cosine similarity computation itself adds another $O(3lmn)$ time complexity to the method, when applied onto all pairs of users and articles, while adding up another $O(nm)$ for the construction of the top lists. In overall for the \emph{Vector Space} model the complexity is $O(lm+ln+3lnm+nm) = O(lnm)$. 
As can be seen, our method again does not exceed the
complexity levels of the classical vector space method.

In our opinion, the large overhead generated by the tag clustering process is not that serious for causing any applicability issues in our method. Such overhead is mainly caused by the fact that the input data used for expessing the distances between the
tags are not derived directly from the user's tagging experience, as
it is the case for other traditional methods, such as \emph{Vector Space},
or the method by Peng et al. In our method instead, clustering is computed upon the semantic similarity of tags, and for that reason clustering data remains constant thereafter. 
Therefore, such cost does not contribute to the computational complexity of the recommendation process.
For that reason it sufficies if applying pre-clustering
once, upon system initialization, and then using the clustering data to any predictions computed thereafter. On the contrary, any approaches based on \emph{Vector Space} model would require re-computation of clusters on a regular basis, as the user data change, resulting to significant overhead in the system.

\section{Conclusions and Future work}
\emph{Annotation Competency} of users has very little been explored in Recommender
Systems. 
In this paper we attempted to explore the potential of using the
information derived from the \emph{Annotation Competency} of users for improving
the prediction accuracy of a \emph{Tag-based} Recommender system.
Such type of systems use alone the tags provided by users for computing personalized item recommendations.
Prior works on tag-based recommendations have indicated that there
was still space for improvement.
Our work is motivated by the need to better understand how users'
annotation works and it provides a new insight on how such knowledge could be incorprorated into the mecanism of producing personalized
recommendations. We introduced a new approach which applies clustering
onto the set of tags that works in succession with our proposed
formula for predicting recommendations. Our formula takes into
account the properties of \emph{Diversity of Concepts} and \emph{Annotation Contribution} we introduced for describing the notion of \emph{Annotation Competency}.
We attempted evaluation on our proposed model using data from
\emph{citeUlike}, a public annotation system for scientific documents.
Our experimentation showed that, if the above two properties are put together, it can help substantially to increase the benefit expressed in terms of
recommenations quality for users. At the same time, the proposed
method was found to be equally computational efficient with other baseline
approaches.

We believe that our work will make significant impact on on-line
Searching and Recommendation services as its simplicity and its low
overhead makes it suitable for such services.
We note the importance of getting a better undestanding of the mechanism of the users' annotation excersize. A wider comparison against
more Tag-based recommendation algorithms is left as future work.
Another important future work is to confirm our conclusions on more
annotation datasets. Investigating our method from the security point of view is also an interesting research direction.

\bibliographystyle{abbrv}
\bibliography{tagging}

\end{document}